# ASSESSING GENDER BIAS IN THE INFORMATION SYSTEMS FIELD: AN ANALYSIS OF THE IMPACT ON CITATIONS

Silvia Masiero, University of Oslo, silvima@ifi.uio.no

Aleksi Aaltonen, Temple University, aleksi@temple.edu

**Abstract:** Gender bias, a systemic and unfair difference in how men and women are treated in a given domain, is widely studied across different academic fields. Yet, there are barely any studies of the phenomenon in the field of academic information systems (IS), which is surprising especially in the light of the proliferation of such studies in the Science, Technology, Mathematics and Technology (STEM) disciplines. To assess potential gender bias in the IS field, this paper outlines a study to estimate the impact of scholarly citations that female IS academics accumulate vis-à-vis their male colleagues. Drawing on a scientometric study of the 7,260 papers published in the most prestigious IS journals (known as the AIS Basket of Eight), our analysis aims to unveil potential bias in the accumulation of citations between genders in the field. We use panel regression to estimate the gendered citations accumulation in the field. By doing so we propose to contribute knowledge on a core dimension of gender bias in academia, which is, so far, almost completely unexplored in the IS field.

**Keywords:** Gender; Information Systems; gender bias; citation impact

## 1. INTRODUCTION

Gender bias is an unfair difference in the way men and women are treated in a particular domain. Psychology has studied gender bias primarily in behavioural terms (Ceci & Williams, 2011; Al-Gazali, 2013), explaining it as a tendency that produces behaviours that penalise women, and favour men, in given contexts. Studies of gender bias have been often formulated in relation to employment, where gender has been found to affect employment opportunities, expectations and career progression across industries (Shen, 2013; Annabi & Lebovitz, 2018). Among these studies, biases are present as recurring patterns: women, research has shown, are comparatively less prone to enter industries that are perceived as gender biased, and more incline to leave such industries (Handley et al., 2015).

Having originated in relation to employment, research on gender bias has diffused to academia, where numerous studies have been conducted to assess gender bias across academic fields. In these studies, gender biases have been found associated with the key dimensions of academic careers including hiring decisions (Reuben et al., 2014), publication quality perceptions (Knobloch-Westerwick et al., 2013), peer review (Helmer et al., 2017), citations (Lariviere et al., 2013), and tenure decisions (Jaschik et al., 2014). Importantly, these studies have often been designed to capture subtle biases, whose silent nature makes it difficult to detect and punish them as outright discrimination (Ceci & Williams, 2011). While covering many different fields, studies of academic gender bias are especially concentrated in Science, Technology, Mathematics and Engineering (STEM) disciplines, where multiple forms of gender bias are found (Handley et al., 2015).

Against this backdrop, we first overviewed studies of gender bias in IS by conducting a literature review of the top journals of the field (known as the Association for Information Systems - AIS Basket of Eight).[1] Drawing on data for all papers ever published in such journals, our review presented a disconcerting picture: out of 7,260 papers for all years, only eight make some reference to gender bias, and out of these only four offer some form of scientometric studies (three focus on journal publications, one on editorial boards). The contrast with the plethora of studies of gender bias in STEM is strong, and greater disconcert emerges from the strong signals of gender bias that emerge in non-scientometric studies of IS (Adam, 2002; Wilson, 2004; Gupta et al., 2019; Winter & Saunders, 2019). Such a contrast between perception of bias and studies of it leads to the question: *Does gender bias exist in the IS field, and if so, how is it manifested?*

To address the question, we focus on research citations that are especially important for academic researchers and can tell about unconscious biases against a gender. Specifically, we aim to estimate the impact of scholarly citations that female IS academics accumulate vis-à-vis their male colleagues. We do so with a study of the AIS Basket of Eight journals: drawing on a database of all papers published in these journals, we plan to study the impact of citations of female vs. male scholars using panel regression and controlling for the length of an academic career, number of publications and network centrality of scholars. The analysis, currently in progress, aims to reveal whether the IS field is biased in favour of male scholars, measuring bias on a dimension that is central to academic careers.

The work-in-progress paper is structured as follows. We first offer a landscape perspective of how gender is dealt with in the IS literature, illuminating the gap left by the absence of scientometric studies of gender bias. We then put forward our research design to study gender bias, describing our dataset and the analysis currently in progress. In the conclusion we outline the expected contributions of this work, which sets to provide one of the first analyses of citation impact of gender in the IS field.

## 2. LITERATURE REVIEW

Our review of the literature aimed at understanding, in the first place, how gender features in top journals of the IS field. To do so, eight keywords (gender, gender bias, gender discrimination, gender inequality, male bias, stereotyping, sexism) were used to search all AIS Basket of Eight journals, resulting in 312 papers that include at least one such keyword in the title, abstract, or keyword list. Both authors then independently coded each paper as "relevant" or "not relevant" to gender-related research, defined as any form of IS research bearing some form of relation to gender. Conflicts in classification between the two authors were resolved by discussing each case individually, which led us to a final set of 86 relevant papers.

We then conducted a thematic literature review of all 86 papers, which led us to recognise three thematic clusters of gender-related research in IS. A first cluster of 55 papers (Type 1) views gender as an explanatory variable, used to research various topics in the IS field such as technology acceptance, use or user behaviour. In 19 of these papers, gender is the main variable in the research, whereas it is a secondary or control variable in the remaining 36 papers. A second cluster of 23 papers (Type 2) deals with gender imbalances in the IT industry, investigating its causes, consequences and interventions taken against it.

While Type 2 papers deal with the IT industry, a third cluster of only eight papers (Type 3) takes issue with gender imbalances in the IS academia, in several cases referring explicitly to gender bias. Three of such papers contain scientometric studies of IS journal publications (Gallivan et al., 2007; Avison et al., 2008; Avison & Myers, 2017) and one of editorial boards (Burgess et al., 2017),

---

[1] https://aisnet.org/general/custom.asp?page=SeniorScholarBasket, accessed 23 March 2021

declaring concern for gender imbalance but without proposing action to address it. The four remaining papers instead all address gender bias in different forms: Adam (2002) presents a critical framework on gender-asymmetric power relations, while Wilson (2004) proposes a framework for critically studying gender in IS research. Among the two most recent publications, Gupta et al. (2019) present evidence of gender bias from a survey of AIS members, whereas Winter and Saunders (2019) show how gender influences the career choices of an IS academic.

Viewing our literature review in the light of gender bias research outside IS, and especially in STEM disciplines, two points are striking. First, the thinness of the body of research engaging gender bias – and calling it by its name – in the IS field is disconcerting: only eight papers out of 7,260 present some engagement with the phenomenon, and only four contain scientometric studies. Second, this paucity clashes with empirical accounts of gender bias in the IS domain, which accounts such as those by Gupta et al. (2019) and Beath et al. (2021) are particularly powerful in revealing. The gap between perceived bias and lack of research on it leads to our research question on gender bias in the IS field.

## 3. RESEARCH DESIGN

As noted above, gender bias can be studied through multiple methods and along different dimensions. For instance, qualitative interviewing can reveal how gender bias is experienced by individual scholars, whereas quantitative studies can assess its prevalence in a field or a journal. In this research, we adopt an econometric approach to estimate the impact of gender on scholarly citations that female IS academics accumulate vis-à-vis their male colleagues. We intend to use panel regression to estimate the impact of gender bias on the citations of papers published by female academics in the field. We chose the dimension of citation impact due to its centrality to academic careers, matched by the availability of usable citation data for all papers in the AIS Basket of Eight journals.

### 3.1. Data collection

To empirically identify gender bias in IS publishing, we use data from the Microsoft Academic Graph (MAG). MAG is the most comprehensive dataset on scientific publishing that attempts to map all academic publications, their authors and citations as a graph (Herrmannova & Knoth, 2016; Wang et al. 2019). We draw from MAG details of all 7,260 publications in the AIS Basket of Eight, including their authors and citations. However, we also need to know the gender of each of the 8,197 authors in the data. We first used the GenderAPI service, a widely utilised service for gender attribution, to assign gender to each author based on first and last name. In our dataset, the service provides a gender attribution (female or male) in 6,104 cases with 95 percent or more confidence. We manually check and correct 2,093 cases in which the attribution is either missing or less than 95 percent confident to ensure reliable gender attribution. As a result of this process, we construct an author-month panel dataset that traces each scholar's publications, citations, and network centrality.

A limitation of the GenderAPI service is its binary attribution of gender, resulting in inability to capture non-binary gender realities (Callis, 2014). Aware of this limitation, we reviewed existing studies of gender bias across fields in order to find services that account for non-binary identities. While such a research has been so far unfruitful, the use of Internet search in combination with GenderAPI (Marrone et al., 2020) helps us mitigate the limits of the service, and input from gender studies research is currently being sought to make sure we take all steps needed to reflect non-binariness.

### 3.2. Data analysis

Our identification strategy is based on the following insight. The pinyin (the official romanisation system for Mandarin language in mainland China) romanisation of Chinese names is generally gender neutral, whereas in most other languages the first name of a person tends to reveal the person's gender. Since there is a substantial number of Chinese scholars in the field of IS, name-related knowledge can be reasonably used to identify the causal impact of gender bias.

Since non-Chinese scholars cannot easily recognise the gender of Chinese academics from their romanised names, there should be no difference in how non-Chinese scholars cite Chinese male and female academics. However, when non-Chinese academics cite non-Chinese scholars they may be implicitly aware of the author's gender and thus show bias against female authors. Any difference between how non-Chinese authors cite other non-Chinese vs. Chinese authors may thus stand for a gender bias, given that we control for a few factors, listed below, that may otherwise account for such differences. These run from the fact that the samples of non-Chinese and Chinese scholars may be different with respect to factors that affect the number of citations a scholar accumulates.

We use time fixed effects to account for changes in the overall citations over the years. On an individual level, we use the length of an academic career, the number of publications, and network centrality to control for the visibility of a scholar in the field. If the differences in citation patterns persist after controlling for these factors, the remaining differences can be attributed to a prejudice against female scholars based on their first name. It is also possible that gender bias is strongest against junior academics and diminishes among senior scholars who have established their presence in the field. Ultimately, if gender bias disappears as the female scholar becomes better known in the field, we may see that senior female scholars receive relatively more citations as their male colleagues, since they have had to work harder to make their name and thus produce higher quality papers.

At the time of writing, we are in the process of checking the 2093 cases in which the GenderAPI attribution is either absent, or assigned with a level of confidence lower than 95 percent. Once this process is complete, we will run the analysis to detect the influence of gender on citation patterns, seeking to understand whether findings of gender-biased citation patterns in fields such as STEM (Lariviere et al., 2013) do, or do not, find confirmation in the IS field.

## 4. EXPECTED CONTRIBUTION AND RELEVANCE TO THE TRACK

The IS field is characterised by substantial anecdotal evidence of gender bias, however accompanied by a striking paucity of scientometric studies of it. The research proposed in this paper aims to contribute to filling this gap, examining gender bias under a dimension, the impact of citations, on which academic careers are substantially predicated. Since IS is the field within which a substantial part of ICT4D research is developed, questions in this study have direct relevance for ICT4D research, offering a method that can be replicated for journals specific of the ICT4D field. By way of example, Davison (2021) offers a study of the AIS Basket of Eight editorial boards vs. the editorial board of the Electronic Journal of Information Systems in Developing Countries, finding significantly higher representation of female editors in the latter rather than in the Basket of Eight journals.

Currently in progress, our work opens several further problems for research. One further question that can build on our study is whether Chinese scholars may present any form of bias towards Chinese female authors, given that they may often recognise the gender from the original (non-transcribed) Chinese name. More questions pertain to the intensity of gender bias in the IS field as related to other fields of academia, and whether its manifestations, strikingly under-investigated so far, mirror those found in fields such as STEM. With our suggested research method, applied to a field where so limited research on gender bias is present, we hope to create more interest for gender bias in IS research, maximising the focus of the field on the topic and setting the basis for its study in ICT4D research works.